# A Conceptual Paper on SERVQUAL-Framework for Assessing Quality of Internet of Things (IoT) Services

Sheikh Muhammad Hizam[1] & Waqas Ahmed[1]

[1] UniKL Business School (UBIS), Universiti of Kuala Lumpur (UniKL), Malaysia

Correspondence: Sheikh Muhammad Hizam, UniKL Business School (UBIS), Universiti of Kuala Lumpur (UniKL), Malaysia.

**Abstract**

Service quality possesses the vital prominence in usability of innovative products and services. As Technological innovation has made the life synchronized and effective, Internet of Things (IoT) is matter of discussion everywhere. From users' perspective, IoT services are always embraced by various system characteristics of security and performance. A service quality model can better present the preference of such technology customers. The study intends to project theoretical model of service quality for Internet of Things (IoT). Based on the existing models of service quality and the literature on internet of things, a framework is proposed to conceptualize and measure service quality for internet of things. This study establishes the IoT-SERVQUAL model with four dimensions (i.e., Privacy, Functionality, Efficiency and Tangibility) of multiple service quality models. These dimensions are essential and inclined towards the users' leaning of IoT services. This paper contributes to research on internet of things services by the development of a comprehensive framework for customers' quality apprehensions. This model will previse the expression of information secrecy concerns of users related with Internet of Things (IoT). This research will advance understanding of service quality in modern day technology and assist firms to devise the fruitful services structure.

**Keywords:** Internet of Things (IoT), privacy, service quality, SERVQUAL

## 1. Introduction

Quality is conceivably the most vital and essential element of business strategy. Organizations struggle on quality, customers look for quality, and markets are renovated by quality. Service quality is considered as single most significant determining element of company's long-term success (Alter, 2010; Ali,et.al 2018). Service quality always has been the influential factor for the successful ventures because the customer's perception towards the company's services possesses the vital significance. Measuring service quality from customer's viewpoint there are several measurement scales formulated. The pioneer model postulated for service quality scaling was SERVQUAL (Parasuraman, Valarie A., & Leonard L., 1988). The model provided the customer perception scaling through various dimensions of service quality. With the time, the nature of services delivered got more complex as compared to merely face-to-face client-company interaction; therefore more service quality models were introduced. Similarly, the evaluation of technology in service delivery made the scenario worthier to expand the quality measurement models. The scenario towards measuring the quality dimension in service sector has been changed since the SERVQUAL model introduced in 1988. Now the digital era is leading the world to a state from physical equipment to virtual & augmented reality, from human-to-human interaction to computer-to-human and computer-to-computer interaction, from manned service counter to digital kiosk desk, that obviously have triggered the aspects of service quality to the variation. This century is period of Industrial Revolution (IR 4.0) where each aspect of service delivery and client perception will be altered. The basic element of IR 4.0 is Internet of Things (IoT) where every equipment and machine under human practice is treated as connected and respondent identity (C.-L. Hsu & Lin, 2018; Audu, 2018).

IoT portrays the theme of forthcoming digital technology, the network of interlinked systems to communicate data in collaboration of convenience and economic advantage (Luthra, Garg, Mangla, & Berwal, 2018). IoT works in every sector like transportation, manufacturing, healthcare, energy, disaster management, government, aeronautics & social media, waste resource efficiencies, environmental governance etc. Industry research reports anticipated huge growth in IoT system around 2020. As 50 billion of IoT-devices with trillions of market value (Weinberg, Milne, Andonova, & Hajjat, 2015). And by 2025, value of internet of things businesses will reach to $11.1 trillion (Manyika, Woetzel, & Dobb, 2015). While in Malaysia the IoT Market would touch the figures of $10 billion by 2025 (MIMOS, 2015). IoT based ETC setup operates by various sensor systems like RFID, Infrared, Microwave, and GPS etc.(Mali, Barge, Kulkarni, Mandhare, & Patil, 2017; Asif, et.al 2018).

The expansion and recognition of sensing and embedding technology with advent of fostering economical wireless chips and tags boosted the Internet of Things (IoT) system that included in various fields of daily life as depicted in the Figure 1, from healthcare to automotive sector etc. The concept of Internet of Things is thriving day by day as total interlinked things (equipment) are escalating to intensifying level of usage (White, Nallur, & Clarke, 2017; Artha, & Mulyana 2018). The huge quantity of equipment serve through various facilities like smart home, smart learning, smart cars, smart cities, smart roads, smart manufacturing etc. (Bellavista, Cardone, Corradi, & Foschini, 2013). And assessing the quality of services in such scenario where the automation prevails without any human intervention either from user or company end, intrigues the elaborative service quality scale.

*1.1 Internet of Things (IoT) and Service Quality*

Individual collaboration is main driver of assessment and evaluation of quality and deliverance of services. The scenario of interaction with client has been changed due to tech-enabled service structure. Beforehand, when handling the service supply, customers had contacted through staff, eager to assist in procedure thoroughly.

```
Wearables                                    Health Care
- Entertainment                              - Remote monitoring
- Fitness                                    - Ambulance telemetry
- Smart watch                                - Drug tracking
- Location and tracking                      - Hospital asset tracking
                                             - Access control
Building and Home Automation                 - Predictive maintenance
- Access control
- Light and temperature control              Smart Manufacturing
- Energy optimization                        - Flow optimization
- Predictive maintenance                     - Real-time inventory
- Connected appliances                       - Asset tracking
                                             - Employee safety
Smart Cities                                 - Predictive maintenance
- Residential e-meters                       - Firmware updates
- Smart street lights
- Pipeline leak detection                    Automotive
- Traffic control                            - Infotainment
- Surveillance cameras                       - Wire replacement
- Centralized and integrated system control  - Telemetry
                                             - Predictive maintenance
                                             - Car to car, and car to infrastructure
```

Figure 1. Application of Internet of Things

Source: Texas Insteument (2019).

Nowadays the evolving influence of innovations, it is noticed that there are plenty exemplars of SST (self-service technology) and automation-based facilities. For instance, online airline check-in, SST checkout lanes in retail shops, fuel pumps and superstores etc. Because of digital arena, now organization can reach to client in every part of globe through online services with less cost and more efficiency. It portrays better deal of how great at considering better opportunities technology able to offer clients with company's perception towards specification of service quality. Service quality is considered as main feature in innovation diffusion because of momentous relationship among client's apparent benefits, service quality, contentment and steadiness towards utilization of electronic facilities. Service convenience and expediency lie in quality of services. (Alsamydai, 2012; Ngala & Mawo 2015; Amogechukwu & Unoma 2017) that determine utilization pattern of users. Perceived service quality is explored as divergence between client's expectations and assessments of service execution. The perception and awareness of customer behavior regarding service quality keeps an essential role in transport studies (Alonso, Barreda, Olio, & Ibeas, 2018)*.* Service quality for IoT based services can be assessed through service quality models measuring the various technological services like SERVQUAL (Parasuraman et al., 1988), E_SERVQUAL (Parasuraman, Zeithaml, & Malhotra, 2005) and SSQUAL (J.-S. C. Lin & Hsieh, 2011) models.

Various approaches in technology sector's service quality postulated to assess services accomplishments. Numerous are acknowledged and proclaimed dimensions as system tractability, accessibility, system reliability, accuracy, response time, comprehensiveness, trustworthiness, functionality, assurance and security etc. SERVQUAL is most famous and its extensions for online or electronic business like E-SERVQUAL, SiteQual, WebQual etc. are commonly used in assessing the service quality in perspective fields (Parasuraman et al., 1988, 2005). Service quality model has positive impact on technology acceptance variables in various sectors of tech-enabled services(Alsamydai, 2016; Boon-itt, 2016; Chang, Pang, Michael Tarn, Liu, & Yen, 2015; Considine & Cormican, 2016; C. S. Lin & Wu, 2002; Majid, Bakri, Shazwani, Junaidi, & Buang, 2014; Sepasgozar, Hawken, Sargolzaei, & Foroozanfa, 2018). While current technologies like Internet of things (IoT), big data, artificial intelligence etc. mulls for service quality scale pertaining the measures of major apprehension of success factor towards customer satisfaction.

For instance, in transportation system, road tolling on expressways was manually controlled and staffed toll plaza used to serve the commuters for paying toll on cash basis service. As technology boosted, the crewed toll plazas replaced with self-service technology by tapping the toll card or inserting cash in machines to pay tolls by drivers. While modern day technology enabled the motorist to mount the RFID tag on their vehicle and pass the toll plaza without stopping and transaction are completed by data i.e., bank or credit/debit card details etc. stored on RFID tag while description of activities of RFID tag later on deliver to mobile phone or portal. This phenomena encapsulating the IoT services aim for open road tolling or gateless tolling (Saad, Abdel-Aty, & Lee, 2018; Aremu, 2018). Similarly, the other IoT services like Smart Home where each equipment of home from Kitchen appliances to drawing room facilities, from parking to surveillance system, all equipped with sensors to synchronize the human life more systematic and managed and this internet of things settings mitigates the human struggles for causal task at home (Kim, Park, & Choi, 2017; Aremu & Ediagbonya 2018). Such disruption from manual to flawless automatic mechanism of service delivery possess the various aspects of customer satisfaction scale, to unveil this will look at the better understanding of customer satisfaction towards modern technology (Haseeb, Abidin, Hye, & Hartani, 2018).

Literature on IoT indicated that studies on IoT were mainly focused upgradation of structural design to facilitate equipment in enhancing direct control usage, boost information assortment and sharing information to one another out of any non-machine interference (Koreshoff, Robertson, & Leong, 2013; Njegovanovic, 2018). But SERVQUAL model for Internet of Things (IoT) services is largely unfamiliar. Because Internet of Things (IoT) operates different from manual service, e-service or self-service technologies and works on automation without any human interaction. Based on this distant nature, the dimensions of IoT-service quality model involve the security, privacy, tangibility, functioning ability and efficiency of system that signal towards the explicit compulsion of SERVQUAL model for IoT based services. This study purposes the four dimensions of IoT based service quality model that comprises of Privacy, Functionality, Efficiency and Tangibility. This paper will undertake the three service quality models i.e., SERVQUAL (Parasuraman et al., 1988), E-SERVQUAL (Parasuraman et al., 2005) and SSQUAL (J.-S. C. Lin & Hsieh, 2011) to propose the dimensions of internet of things (IoT) services. The proposed service dimension would be active player in measuring the customer expectations from service providing organization to induce customer retention and satisfaction.

## 2. Literature Review

*2.1 Service Quality*

Service is appliance of particular capabilities, via distinct procedures to assist individuals (Vargo & Lusch, 2008). While service quality was delineated like "An overall evaluation or attitude towards the comprehensive excellence of service" (Parasuraman, Zeithaml, & Berry, 1985). Far along, Alter (2010) tracked the novel technique that described service quality in a way like "The approach which is based on the predictable attitude and depends on the assumptions about the connection between service quality assessment and subsequent behavior which are not supported by the efficient body of research conclusions about consumer's behavior. The attitude-based method also causes inferences to be equipped with reference to what facets of service delivery control the attitudes". Service quality phenomenon is significant and pondering subject matter in service providing organization (Jermsittiparsert et al., 2016). Now days in era of digital economy the service quality is matter of discussion in internet of things (IoT) services (Bello & Zeadally, 2017; Nkiru, Sidi & Abomeh 2018). Nature of services and customer's perceptions included the new aspects such as elusiveness, delicateness and inextricable made service quality measures complex as compared to tangible services (Parasuraman et al., 2005; Almasi & Khorasgani 2018).

Service distribution process impacts consumer's perception by staff client handling (Ramseook-Munhurrun, Lukea-Bhiwajee, & Naidoo, 2010). Parasuraman et al. (1988) described service quality as "a global judgment or attitude relating to the overall excellence or superiority of the service" and abstracted the consumer's assessment of service quality in accordance with Oliver's (1980 cited by Shabbir, Malik, and Janjua, (2017) disconfirmation paradigm. This is the chasm in service expectation and perception equates. Moreover, SERVQUAL postulated the mainly measure of service quality with numerous elements that denoted the customer trust, customer caring, service equipment aesthetics, and compassion.

## 3. SERVQUAL

Service delivery receives recognition and influence by its level of quality. It includes diverse features of quality like reliability of service, receptiveness of service provider, empathy, reassurance and tangibility (Parasuraman et al., 1988). SERVQUAL model was theorized to establish scale of quality of services provided by organization and governments. Parasuraman et al. (1988) abstracted service quality in five dimensions measurement concept that comprised of (1) Reliability: the extent to which a guaranteed service is performed reliably and precisely", (2) Responsiveness: the scale to which service sources are inclined to assist clients and deliver timely service, (3) Assurance: the degree of service facilitators are conversant, well-mannered, and capable to invigorate trust, (4) Empathy: extent to which clients are proffered caring and personalized consideration, and (5) Tangibility: the extent to which physical resources, equipment, and appearance of staffs or personnel are ample.

Recently in Italy, customers' opinion (between positive and negative) about product quality analyzed by SERVQUAL (Palese & Usai, 2018). *Moreover, SERVQUAL assessment explored dimensions possesses various dispersals in grading and strength*. Consumer's preference drives the organizational services quality development and upgradation. As an instance, receptiveness of vendors from clients' point of view entails the service exposure at unduly highly priority than some more matter (Palese & Usai, 2018). This model has various extension in field of modern day technology like E-SERVQUAL, WebQual, SiteQual, IRSQ, eTailQ, PeSQ, SSTQ (J. S. C. Lin & Hsieh, 2011).

*3.1 E-SERVQUAL*

Various studies directed that individual to innovation collaboration pointed towards consumer evaluation of modern technology as a distinctive practice (Gao & Bai, 2014). After penetration of internet in daily usage, services also became electronic as e-commerce, e-learning etc. Electronic service is defined as starrer in digital market service (Rust & Lemon, 2001). Later on, as business shifted to digital services for their customers like websites or web portals, the aspects of services quality updated to level of privacy of clients' data on websites, system availability, and efficiency of websites and fulfilment of service purpose. This model called as E-SERVQUAL or E-SQ (Parasuraman et al., 2005; Almeqdadi, 2018). This model was basically formulated to rank the interaction of client towards the websites of businesses or governments or educational institutions in relation to the offered services. Conferring Parasuraman et al. (2005) E-SERVQUAL Scale, comprising of 22 items on four components encapsulating **1.) Efficiency:** The convenience and speed of retrieval with utilizing website. **2.) Fulfillment:** The degree to which the website's vows towards order distribution and item accessibility are fulfilled. **3.) System availability:** The accurate technical performance of the site. **4.) Privacy:** The extent to which website is secure and shields consumer's data.

E-SERVQUAL hypothesized the service quality scaling in various sectors of business. Customer's privacy in online shopping or sharing personal information considered as basic scale of better service quality. Website performance and it's easy to use frame for all clients rank the quality criteria. System availability and fulfilment of task also measuring factor of electronic service quality. It has widely used in measuring the level of service quality for websites and mobile applications for instance online support system by government for employees in Spain, (Janita & Miranda, 2018) group-purchase behavior measures in Taiwan from social media website like Facebook (S. Hsu, Qing, Wang, & Hsieh, 2018), online taxi service (Alonso et al., 2018), healthcare (Hee, Kim, & Won, 2016) and banking sector (Alsamydai, 2016; Rostami, Amir Khani, & Soltani, 2016).

*3.2 SSQUAL*

Customers interacted with automation aspect of services after online services. Clients collaboration with company's services has innovatively boosted due to self-service technology. (J. S. C. Lin & Hsieh, 2011). Customers' exposure towards service provider varies in a way like service provided by staff or self-service coproduced by clients themselves. Increasing workforce expenditure have invigorated businesses to discover additional service alternatives to let consumers carry out services on self-basis. Innovation development has heightened service dispensing with modernized background, permitting businesses to practice various autonomous facilitation tools to enhance the client's partaking. Self-service technologies are installed to supersede the crew based amenities from super markets' point of sales to daily banking functions (Lin & Hsieh, 2011). Earlier digital facilities quality characteristics were based on cybernetic aspect like web-portal whereas self-service tools progressed to manifold services network coproduced by customers. Therefore Service quality model for Self-Service Technology (SSQUAL) was introduced by Lin & Hsieh (2011) that consists of seven dimensions of quality measures that include **1) Functionality** exemplifies the practicable attributes of self-service system, entailing degrees of responsive, reliable, and convenient usability **2) Enjoyment** illustrates insights of delight or satisfaction experienced through SST service distribution, **3) Security/Privacy** portrays apparent security from infringement, deception, and theft of personal data. **4) Assurance** depicts trust level because of company's repute with competency of Self-service technology. **5) Design** denotes total depiction or outline of self-service technology setup. **6) Convenience** describes approachability in self-service technology mechanism. **7) Customization** validates the scenario where self-service system structure possesses capabilities of restructuring and reforming in order to fulfil the client requirement. Applications of SSQUAL are dispersed in every field of business and appeared as an improved tool compare to other service quality measure instrument ( Ahmed, Abdul Majid, Mohd Zin, Phulpoto & Umrani, 2016; Boon-itt, 2016; Considine & Cormican, 2016; Demoulin & Djelassi, 2016; Einasto, 2014; Siah, Fam, Prastyo, Yanto, & Fam, 2018).

*3.3 IOT-SERVQUAL*

From typical services by crewed setup to online websites or e-portals and then to self-serving automatic machine, all frames are engulfed by the scales for gauging service quality**.** Literature pertained validity of such models (S. Hsu et al., 2018; Janita & Miranda, 2018; Palese & Usai, 2018; S á, Roch, & Cota, 2016; Sepasgozar et al., 2018; Shahid Iqbal, Ul Hassan, & Habibah, 2018). The IoT is an interconnected setup of electronic things entailing various devices motorized or digitized base, humans or animals that are categorized through distinct identification pattern, the UIDs, to communicate the information throughout system (Rouse, 2016). Internet of things (IoT) involves automatic services with electronic interaction of website or mobile application by physical equipment of system. That point towards the scenario of measuring the service quality for IoT services involves the above mentioned three quality scale models. Dimensions of IoT service quality models resides in previously discussed models as the newly proposed quality model dimensions mostly lied in preceding established models (J.-S. C. Lin & Hsieh, 2011; Parasuraman et al., 2005). The dimensions of these three models are depicted in Table 1.

Table 1. Service quality models

| SERVQUAL | E-SERVQUAL | SSQUAL |
| --- | --- | --- |
| **(Parasuraman et al., 1988)** | **(Parasuraman et al., 2005)** | **(J.-S. C. Lin & Hsieh, 2011)** |
| **Tangibility** | Efficiency | Functionality |
| **Assurance** | System Availability | Enjoyment |
| **Reliability** | Fulfilment | Security/Privacy |
| **Empathy** | Privacy | Assurance |
| **Responsiveness** | | Design |
| | | Convenience |
| | | Customization |

Though SERVQUAL (Parasuraman et al., 1988) works as prevalent theme to evaluate the service quality in client interaction with businesses. Studies have explored that consumers' quality valued aspects, in novel innovation services are different from conventional customer dealings therefore E-SERVQUAL model was presented that covered the websites, online portals services offered by business or government to client and citizens (Parasuraman et al., 2005). The SSQUAL towards self-service technology developed by Lin & Hsieh, (2011), extends ahead of cyberspace as service distribution network. It entails 20 items that assess perceived services quality, irrespective alternate dissemination network. But the advancement in technology and innovation on daily basis requires the parallel service quality scale. For instance, IoT based services are unable to be measured properly through previous model as these pertaining either website quality or kiosk self-serving equipment quality measures. But in this context Internet of things-based services literature like connected vehicles, smart homes etc. for such model is largely unknown.

The IoT-SERVQUAL is proposed to cater the measurement of IoT services and it includes the measurement scale from three above discussed models of SERVQUAL, E-SERVQUAL and SSQUAL. IoT based service quality model incorporated the items from preceding models as per Table 1 like Privacy that point out realized protection in state of interruption, deception, and private data breach, is adopted from E-SERVQUAL and SSQUAL. Functionality that depicts the reliability, usage convenience and responsiveness is engaged from SSQUAL. Tangibility, the services apparatus outlook and visible features from SERVQUAL and Efficiency the speed of process and its related aspect is pledged from E-SERVQUAL.

3.3.1 Privacy

Privacy implies when business keeps the client data secure and only with their consent, disseminate to other business activities. This involves reassurance of classified piece of data communicating from consumer to organization and vice versa. Furthermore, it assures protected network connection through graphical depiction (Collier & Bienstock, 2003). In literature contemplating the internet of things based customers' point of view, matters of security, confidentiality and attributes of devices play vital role towards system acceptance and in shaping the buying behavior (Lu, Papagiannidis, & Alamanos, 2018). Internet of Things services mechanism is backed by various associated procedures of generating, acquiring, broadcasting and analyzing the data. Internet of things has no value without data. Undeniably, Internet of things is merely discussion of data, especially regarding customers. Being as only tool of a business enterprise in acquiring and employing, it is advantageous in deducing the individuals' date of birth, salary, activity on website and social media. However, the information like humans' daily activities such as nutritional plans, health statistics etc. forcibly deduced by interconnected setup of devices is utterly different phenomenon. Internet of Things keeps the total record of human activities and nature of his living in shape of piece of information and this information is circulated from device to device like an individual being the data (Weinberg et al., 2015). Therefore, admiration of client's discretion possesses the vital concern in exposure towards internet of things service. To get benefited by convenient related amenities of Internet of Things, the customers will compromise and intrigues for inadequacies in keeping their data secure (Wottrich, van Reijmersdal, & Smit, 2018).

Being the main theme of assurance and affinity mechanism, the secrecy and confidentiality are point of focus. Diffusion of innovation in internet of things service can heighten the level of interaction with high cost of privacy stakes. Numerous global organizations like Apple, Sony and Target etc. had faced the prevalent intrusion into their systems. That had revealed and compromised the useful information of citizens like their official identity numbers towards government interaction, business policies etc. that resulted in high monetary and privacy loss. Internet of Things setup links the entities to internet via sensors. Stealing the information from certain databank is an issue but intruding someone's digital network and its manipulation is devastating. For instance, various details and stories pointed towards meddling sensitivity of vehicles system that caused getting the control of motorcar system, (Reindl, 2018). Same way the security breach in healthcare IoT setup would cause the numerous shapes of threats to individuals' life matters. Discretion is matter of discussion while implementation of Internet of Things services. And the data sharing on sensors, data usage through mobile application and website activities always had the concerns for the customers (Ding, Jiang, & Su, 2018; Feng & Xie, 2018; M. Turri, J. Smith, & W. Kopp, 2017; Wottrich et al., 2018). For successful service delivery, organizations need to build their facilities structure with higher measures of privacy and security assurance as these are the most concerning aspect of IoT services acceptance and customer satisfaction.

3.3.2 Efficiency

The efficiency is assessed as the time it takes various users to accomplish the task, the cost of retrieving the e-service and the quality of the offered service (Parasuraman et al., 2005). The Internet of Things (IoT) is expected to contribute distinct and complicated accomplishment in days ahead that will more support in efficient service delivery (Al-Shammari, Lawey, El-Gorashi, & Elmirghani, 2018). IoT services operate through certain mobile application or website, and the efficient service delivery is based on online interface for service usage. It involves how easy and speedy way the information can retrieve, the payment can process, function can render. Mobile application or Website should effortless to use, properly designed and necessitate a least amount of information to be punched by the customer are the basics of efficient service delivery. It also involves in financial aspect, the speed of service for instance clearing, depositing, query, fetching information, money transfer, response, speedy transaction and verify out with minimum time. Efficiency had proven to be the most influential service quality dimension in websites quality measurement (Parasuraman et al., 2005). Layout of mobile app in Healthcare IoT system, the application efficiency should pertain the speedy way response of the varying symptoms in body as per sensors response. In automobile system, the connected car system should respond the drivers about the certain vehicles impairments, road situation and traffic updates. The system is efficient to facilitate the customer with timely responding the information of its functions so the client can take decision accordingly (Khairi Majid, Bakri, Shazwani Laila Junaidi, & Roslan Buang, 2014).

3.3.3 Functionality

Functionality, as a system's potential, is anticipated to give users what they need and to assist the organization meet its strategic goals. In information system research, the extent to which the information system functionalities meet the needs of the task regulates system effectiveness for that task (Barkhi, Belanger, & Hicks, 2008). According to Grönroos (1984) quality perceived by customers is based on two approaches, technical and functional. Technical quality is ultimate result what customers acquire while functionality is the process, the way; the procedure customer gets the service delivery. This functional aspect of service quality contains particular nature (Fiala, 2012). To assess the service quality for automated self-service technology, functionality explored the level of easiness and convenience to use the technology, how much this automated system is reliable, and possessing the alertness towards the user (J.-S. C. Lin & Hsieh, 2011). IoT devices skilled in managing manifold sensor based stimulating events that will become foundation layer of impending internet of things system regarding urbanization (Weinberg et al., 2015). IoT services functions through website or mobile app, the ease of use in functionality points to the extent at which the user can operate the system in hassle-free way. Reliability depicts the ability of system to properly accomplishment of the corresponding task. For instance, in smart home, the user required to properly control each section of house through app or user-interface setup. User satisfaction depends on the convenient functionality and task fulfilment of system. It has been widely practiced in automated technology (Bhattacharya, Gulla, & Gupta, 2012; Boon-itt, 2016; Considine & Cormican, 2016; Sá, Rocha, & Pérez, 2015).

3.3.4 Tangibility

Tangibility is the extent to which physical resources, equipment, and appearance of staffs or personnel in adequate manner (Parasuraman et al., 1988). Tangibility for digital services based on design of websites, contents of site, visualization etc. (Moon, 2013). The concept of tangibility towards internet of things seems unusual but interfacing and collaborating with such devices does not remain an uniqueness idea anymore as it initiated already in 2010, Kranz, Holleis and Schmidt (2010) demonstrated several illustrations in this field. Mayer, Tschofen, Dey and Mattern (2014) suggested the permutation mechanism of semantic communication primitives to various IoT entity attributes. That aims to support both web graphical user interface (GUI) and quantifiable interfaces, for example the tangible knobs. According to Angelini et al. (2018) the concept of tangibility in internet of things was introduced by Sarah Gallacher (ETIS, 2017) to stimulate the change in layout of physical interfaces with Internet of Things. Tangibility of IoT services with sensors technology and mobile phone app exemplified by various practical tools. TANGERINE was a system for maintaining cooperative interactions on tabletops where customers could transfer data to perform certain activities via client-centered constructed Internet of Things based tangible dice (Baraldi et al., 2008). The couple of articles introduced distinct such application setups: for collaborative traffic management. (Lebrun, Adam, Mandiau, & Kolski, 2015) and in food business for inferring recipes (Lebrun, Lepreux, Haudegond, Kolski, & Mandiau, 2014). De la Guía et al. (2015) explored the mechanism aiming the acceleration of smart home supervision via tangible Near-Field Communication (NFC) cards linked with numerous matters of actions. The smooth interaction of IoT objects by considering the characteristics of tangibility of application like mobile app or website layout, design, aesthetic manner with device layout like device design, device aesthetic, visual appealing of system, gadget design synchronization with nature of provided service etc. affect the user satisfaction. Similarly Moon (2013) explored that the tangibility of digital services boosts the client satisfaction and creates elicit positive expression of provided facilities.

## 4. Conclusion

The future belongs to Internet of Things (IoT) services. The technology diffusion with rapid growth intrigues for customer acceptance and satisfaction. Embedding everything with sensors (mostly RFID) to interact with human and other object in order to complete service delivery, service quality will be major determinant in smooth adoption in this scenario. The nature of IoT-service has evolved through manifold features as from physical interaction to online presence with automatic service delivery. The proposed dimensions of service quality would support to identify the customer satisfaction measurement for Industrial Revolution 4.0 (IR4.0) based services. This conceptual expression will allocate businesses to gain a better comprehension into what customer's judge in the assessment of IoT-service quality. This paradigm for IoT-service quality is conceptual and will require empirical backing to assess the proposed dimensions that customers use in evaluating IoT-service quality. The research for IoT-service quality is in the initial stages with numerous opportunities for advance research to inflate the knowledge in this area.